# The correlation coefficient between citation metrics and winning a Nobel or Abel Prize

M.V. Simkin

Department of Electrical and Computer Engineering, University of California, Los Angeles, CA 90095-1594

mikhailsimkin@hotmail.com

**Abstract.** Computing such correlation coefficient would be straightforward had we had available the rankings given by the prize committee to all scientists in the pool. In reality we only have citation rankings for all scientists. This means, however, that we have the ordinal rankings of the prize winners with regard to citation metrics. I use maximum likelihood method to infer the most probable correlation coefficient to produce the observed pattern of ordinal ranks of the prize winners. I get the correlation coefficients of 0.47 and 0.59 between the composite citation indicator and getting Abel Prize and Fields Medal, respectively. The correlation coefficient between getting a Nobel Prize and the Q-factor is 0.65. These coefficients are of the same magnitude as the correlation coefficient between Elo ratings of the chess players and their popularity measured as numbers of webpages mentioning the players.

## Introduction

There are adverse views on the usefulness of citation index for research rating. Ranging from claiming that it perfectly reflects research value (1) to calling it a "cuckoo in a nest" (2). Mathematical models of citation copying can account for highly cited papers using ordinary law of chances (3).

Recently Kosmulski (4) studied how citation numbers match with Nobel Prize winning. He checked the list of 100,000 most cited scientists compiled by Ioannidis et al (5). Out of 97 recent Nobel prize winners only 90 were on that list. Just 45 were among top 6000. I checked the same data set (5) and discovered that it includes only 6 out of 22 Abel Prize winners. Andrew Wiles who proved Fermat theorem is prominently missing. Only 22 of 60 Fields Medal recipients are on that list. A skeptical reader may argue that the culprit is that mathematicians are far less cited than the other scientists. While formally true, practically the argument is wrong. The list of 100,000 most cited scientists (5) contains 1.45% of all scientists and 0.93% of all mathematicians. So, mathematicians are only slightly underrepresented. In fact, the list includes almost 900 of them.

Sinatra et al (6) introduced a new citation-based metric which they called a Q-factor. I will describe it in more detail in the Model section. They computed this parameter for almost three thousand physicists. They presented their results in the form of ROC curves (described in the Discussion section) which are not straightforward to interpret. But if you carefully analyze their plots, you discover that out of top 25 Q-factor scientist only 5 are Nobel Prize winners. And the Q-factor rank of the lowliest of the 25 Nobel Prize winners in the sample is over 700.

We see that agreement between the opinion of Nobel/Abel/Fields committees and citation rankings is far from perfect. At the same time, they are not fully independent. In statistics they use a correlation coefficient to describe such relations. It is far more convenient to use a single number rather than state how many Nobel winners are among top X cited scientists and how many are among top Y. The correlation coefficient would be straightforward to compute had we had the rating given by the Nobel committee to all scientists in the pool. Without having those ratings, the task is tricky but not impossible. We do know ordinal ranks of the

Nobel laureates in the list ordered according to citation numbers. And can infer what correlation coefficient was most likely to produce such pattern.

## Model and Results

I will start with describing the Q-model of Sinatra et al (6). It states that the impact of a paper (they use the number of citations a paper gets during first 10 years since publication and denote it $c_{10}$) is given by

$$c_{10} = QP \tag{1}$$

Here $P$ is a lognormally distributed random variable and $Q$ is fixed for each scientist but lognormally distributed among different scientists. The authors with high $Q$ have many highly cited papers. According to (6) the $P$-factor describes the role of randomness or luck and $Q$-factor describes talent or individual ability.

As Wiles does not enter top 100,000 most cited scientists, $Q$ surely is not a mathematical talent. As some Nobel/Abel prize winners are also highly cited it is not entirely unrelated to it. I propose:

$$Q = TR \tag{2}$$

Here $T$ is the true talent or achievement, accurately assessed by Nobel or Abel committee. And $R$ is a talent/achievement unrelated factor. It can be a completely random factor, or it can be a marketing talent. I will not dwell upon it, just state that the factor is uncorrelated with scientific talent. It follows from Eq. (2) that

$$log(Q) = log(T) + log(R) \tag{3}$$

Now $log(Q)$ is normally distributed, as they found empirically in (6). I will hypothesize that $log(T)$ and $log(R)$ are also normally distributed.

A standard way to get two normally distributed (with zero mean and unit variance) variables $x$ and $y$ with Pearson correlation $r$ is:

$$y = rx + \sqrt{1-r^2}\,s \tag{4}$$

Here $s$ is a normally distributed variable (with zero mean and unit variance) uncorrelated with $x$. We may think of $x$, $y$, and $s$ as normalized to have a zero mean and unit variance $log(T)$, $log(Q)$, and $log(R)$ respectively.

Eq.(4) leads to the following joint probability distribution of $x$ and $y$ (7):

$$p(x,y) = \frac{1}{2\pi\sqrt{1-r^2}} e^{-\frac{1}{2(1-r^2)}(x^2 - 2rxy + y^2)} \tag{5}$$

Now Nobel prize winners have the highest $x$. The distribution of $y$ for Nobel prize winners is a conditional probability distribution with the condition $x > x_c$. A calculation gives:

$$p(y|x>x_c) = \frac{1}{2\pi} e^{-\frac{1}{2}y^2} \int_{\frac{x_c - ry}{\sqrt{1-r^2}}}^{\infty} e^{-\frac{1}{2}t^2} dt \Big/ \sqrt{\frac{1}{2\pi}} \int_{x_c}^{\infty} e^{-\frac{1}{2}t^2} dt \tag{6}$$

In Ref. (6) they had 25 Nobel winners in a sample of 2,887 scientists[1]. Thus, being a Nobel winner means being at 0.991 percentile level talentwise. Using inverse cumulative normal distribution, we get $x_c \cong 2.38$. This figure should be substituted into Eq.(6). Next, we compute $y_k$ based on Q-factor rating for every Nobel winner $k$ in the sample. If, for example, Nobel winner number 1 has a Q-factor rating 4, this puts him at 0.999 percentile level. Using inverse cumulative normal distribution, we get $y_1 \cong 3.00$. Next, we maximize the likelihood function

$$L = \prod_{k=1}^{N} p(y_k | x > x_c) \qquad (7)$$

with regard to correlation coefficient $r$ (it enters via $p(y|x > x_c)$, which is given by Eq.(6). )

One can easily do numerical maximization of Eq. (7) in Excel which has built in cumulative normal distribution function and its invers. It gives $r \cong 0.65$. By numerically computing $L$ for a range of $r$ and using Bayesian inference we get that with 95% probability $0.54 < r < 0.74$. This is the inferred correlation between the Q-factor and the rating of the scientists by the Nobel committee.

An alternative approach is to compute the expected fraction of Nobel winners among the top $M$ Q-factor scientists. This will be

$$p(y > y_c | x > x_c) = \int_{y_c}^{\infty} p(y | x > x_c) dy \qquad (8)$$

Here $p(y|x > x_c)$ is given by Eq.(6) and $y_c$ is determined by $M$. A natural choice is to take $M$ equal to the number of Nobel winners in the sample. In that case we get $y_c = x_c$. In the sample in question there are 5 Nobel winners among top 25 Q-factor scientists. The most likely correlation coefficient to produce this result is the one that gives $p(y > x_c | x > x_c) = 0.2$. We should numerically solve this for $r$ using Eqs. (6) and (8). The result is $r \cong 0.63$, which is close to the previous estimate.

Yet another approach is a Monte-Carlo simulation. This should answer the concerns of finite sample size, since the previous results are precise only in the limit of an infinite sample. I produced 2,887 pairs of normally distributed random variables and made them correlated using Eq.(1). I ran the simulation for different values of $r$ with a step of 0.01 using 10,000 samples for each value of $r$. I checked how many times exactly 5 out of top 25 values of x were associated with one of the top 25 values of y. The maximum was at $r = 0.63$ where 2,194 out of 10,000 samples produced this pattern.

Now we proceed to compute the correlation between the composite citation indicator of Ioannidis et al (5) and Abel prize. They compute this composite indicator by summing six different terms[2]. Here we cannot say for sure that the index is normally distributed since we only have the data for top few percent of the scientists. But the data in the available range is consistent with normal distribution. So we will use the same method.

The database of 100,000 most cited scientists includes 898 mathematicians. The total number of mathematicians counted by Ioannidis et al is 96,619. Unlike the preceding case we have no idea of the citation ranking of the most of the Abel winners since they did not enter the list. However, we can use the approach that led to Eq.(8) in the previous case. Being one of 22 Abel winners out of 96,619 mathematicians puts one at 99.98 percentile level, so $x_c \cong 3.51$. Being one out of 898 most cited puts one at 99% level so $y_c \cong 2.35$. Since 6 Abel winners enter the most cited list we should get $p(y > y_c | x > x_c) = 6/22$. Solving this for the correlation coefficient gives $r \cong 0.48$.

---

[1] The 2,887 number is from chapter S1.3 of (6). The 25 Nobel winners number I inferred from Fig.S45 after extracting data using plot digitizer (8). From the same figure I extracted approximate ordinal rankings of the Nobel winners in the sample.
[2] The first term is the logarithm of the total number of citations normalized by the maximum value of this parameter in the pool. So that the scientist with the maximum total citation number has this term equal to 1. The second term is the logarithm of the h-index normalized the same way. The other four terms include similarly normalized coauthorship-adjusted h-index, number of citations to papers as single author, number of citations to papers as single or first author, number of citations to papers as single, first, or last author. They take care to exclude self-citations when computing composite indicator.

The above approach did not use all available information since for some Abel winners we do know their precise citation rankings. So we should maximize the following likelihood function:

$$L = \prod_{k=1}^{N_1} p(y_k | x > x_c) \times \left(1 - p(y > y_c | x > x_c)\right)^{N_2} \tag{9}$$

Here $N_1$ is the number of Abel winners which enter the top cited list and $N_2$ is the number of those who do not. In our case we have $N_1 = 6$, $N_2 = 16$. Numerical maximization of Eq.(9) gives $r \cong 0.47$.

In the case of Fields Medal we have $N_1 = 22$, $N_2 = 38$. Numerical maximization of Eq.(9) gives $r \cong 0.59$.

## Discussion

As we see from the previous chapter, the correlation between getting prizes and citation measures is medium to high. However, in an earlier commentary (9) the present author suggested that it is low. I came to that result while analyzing receiver operating characteristic (ROC) curves used in Ref. (6) to describe the relation between the Q-factor and Nobel Prize winning. They are produced the following way. One selects top *M* Q-factor scientists and counts how many Nobel winners and other scientists got in the sieve. The ratio of the number of Nobel winners in the sieve to their total number in the sample is true positive rate (TPR). The ratio of non-winners in the sieve to their total number in the sample is the false positive rate (FPR). By plotting TPR vs FPR for all values of *M* one gets an ROC curve.

I proposed (9) an alternative way to analyze the data by computing a correlation coefficient. To each scientist we can attribute a Q-number which is 1 if they get in the sieve at given rank threshold and 0 otherwise. And an N-number which is 1 if the scientist got a Nobel Prize and 0 otherwise. Now we can compute a Pearson correlation coefficient between these vectors. A particularly simple result we get for the case when *M* equals the number of Nobel winners in the sample:

$$r = TPR - FPR \tag{10}$$

It is certainly some correlation coefficient. And it equals unity in the case of a perfect match when $TPR = 1$ and $FPR = 0$. However, when I compute the RHS of Eq.(10) using theory presented in the previous chapter I get a result very different from $r$.

Interestingly another way of producing correlated variables, when instead of Eq.(4) we use

$$y = \begin{cases} \text{with probability } r: & x \\ \text{with probability } 1-r: & s \end{cases} \tag{11}$$

produces almost a match to Eq.(10) with small corrections of the order of the square of the relative fraction of Nobel winners in the sample. Albeit Eq.(11) is a rather unnatural way of introducing correlation. So results obtained using Eq.(4) must be closer to reality.

The inferred correlation coefficients between prize winning and citation metrics are of the same magnitude as the correlation coefficient of $r \cong 0.61$ between Elo ratings of the chess players and their popularity measured as numbers of webpages mentioning the players (10). Or the correlation coefficient of $r \cong 0.72$ between the number of victories achieved by fighter-pilot aces and numbers of webpages mentioning the aces (11). This means that ranking scientists by citation numbers is as good or as bad as judging chess players strength or fighter-pilots victory scores based on their fame.

If we substitute the highest correlation coefficient that we got in this study $r \cong 0.65$ into Equation (4) we get that the pre-factor before talent or achievement unrelated random variable $\sqrt{1-r^2} \cong 0.76$ which is about 17% bigger than the pre-factor before talent. This means that talent/achievement unrelated things are mostly responsible for citation metrics like Q-factor.